\documentclass[aps,prl,reprint,superscriptaddress,showpacs]{revtex4-1}
\usepackage{graphicx}

\usepackage{color}
\usepackage{ulem}

\begin{document}

\title{
Copper Doping of BaNi$_{2}$As$_{2}$: Giant Phonon Softening and Superconductivity Enhancement
}

\author{Kazutaka Kudo}
\email{kudo@science.okayama-u.ac.jp}
\author{Masaya Takasuga}
\author{Minoru Nohara}
\email{nohara@science.okayama-u.ac.jp}
\affiliation{Research Institute for Interdisciplinary Science, Okayama University, Okayama 700-8530, Japan}

\date{\today}

\begin{abstract}
The effects of copper doping on the structural and superconducting phase transitions of Ba(Ni$_{1-x}$Cu$_{x}$)$_{2}$As$_{2}$ were studied by examining the resistivity, magnetic susceptibility, and specific heat. 
We found an abrupt increase in the superconducting transition temperature $T_{c}$ from 0.6 K in the triclinic phase with less copper ($x$ $\leq$ 0.16) to 2.5--3.2 K in the tetragonal phase with more copper ($x$ $>$ 0.16). 
The specific-heat data suggested that doping-induced phonon softening was responsible for the enhanced superconductivity in the tetragonal phase. 
All of these observations exhibited striking similarities to those observed in the phosphorus doping of BaNi$_{2}$(As$_{1-x}$P$_{x}$)$_{2}$ [K. Kudo {\it et al.}, Phys. Rev. Lett. {\bf 109}, 097002 (2012).], which markedly contrast the behavior of phosphorus and copper doping of the iron-based superconductor BaFe$_{2}$As$_{2}$. 
\end{abstract}


\maketitle

Nickel-based 122-type compounds, such as BaNi$_2$As$_2$ \cite{ronning,kurita,sefat,subedi,shein,chen,kudo1,yamakawa}, 
SrNi$_{2}$As$_{2}$ \cite{bauer}, BaNi$_{2}$P$_{2}$ \cite{mine,tomioka}, SrNi$_{2}$P$_{2}$ \cite{keimes,ronning2,kurita2,kudo2,hlukhyy1}, 
BaNi$_{2}$Ge$_{2}$ \cite{hlukhyy2,hirai}, and SrNi$_{2}$Ge$_{2}$ \cite{hung}
are nonmagnetic analogs of iron-based superconductors. Some of these compounds have provided opportunities for investigating the interplay between structural instability and superconductivity upon chemical doping \cite{kudo1,hirai,kudo2,hlukhyy1}.

The prototype BaNi$_{2}$As$_{2}$ compound crystallizes in a tetragonal ThCr$_{2}$Si$_{2}$-type structure (space group $I4/mmm$, $D_{4h}^{17}$, No.~139) and exhibits a structural phase transition from its tetragonal phase to a triclinic phase ($P\bar{1}$, $C_{i}^{1}$, No.~ 2) at $\simeq$ 130 K, below which alternate Ni-Ni bonds are formed in the Ni square lattice \cite{sefat}. 
In this triclinic phase, weak-coupling Bardeen--Cooper--Schrieffer (BCS) superconductivity emerges at 0.7 K \cite{ronning,kurita}. 
The superconducting transition temperature $T_{c}$ abruptly increases from 0.7 to 3.3 K upon phosphorus doping of BaNi$_{2}$As$_{2}$ \cite{kudo1}. 
The enhanced superconductivity, which is a strong-coupling type, is accompanied by the triclinic-to-tetragonal phase transition and pronounced phonon softening that is induced by phosphorus doping at $x$ = 0.067 in BaNi$_{2}$(As$_{1-x}$P$_{x}$)$_{2}$ \cite{kudo1}. 
Phosphorous and arsenic have the same number of valence electrons, but phosphorus has a smaller ionic radius than arsenic. Thus, phosphorus doping was anticipated to produce chemical pressure in BaNi$_{2}$(As$_{1-x}$P$_{x}$)$_{2}$ \cite{kudo1}, activating a mechanism similar to the pressure-enhanced superconductivity of elemental tellurium at the rhombohedral $\beta$-Po to bcc phase transition \cite{mauri,rudin,suzuki}.

For iron-based superconductors, the doping of various chemical elements has been examined in order to demonstrate the suppression of antiferromagnetic and structural phase transitions and the subsequent appearance of superconductivity \cite{hosono}. 
In the prototype BaFe$_{2}$As$_{2}$ compound, potassium doping resulted in a maximum value of $T_{c}$ = 38 K in Ba$_{1-x}$K$_{x}$Fe$_{2}$As$_{2}$ \cite{rotter}, while phosphorus doping resulted in a maximum value of $T_{c}$ = 31 K in BaFe$_{2}$(As$_{1-x}$P$_{x}$)$_{2}$ \cite{kasahara,jiang}. 
Interestingly, transition-metal doping resulted in superconductivity with maximum transition temperatures of $T_{c}$ = 23 and 20.5 K for Ba(Fe$_{1-x}$Co$_{x}$)$_{2}$As$_{2}$ \cite{sefat2,ni1} and Ba(Fe$_{1-x}$Ni$_{x}$)$_{2}$As$_{2}$ \cite{li,ni2}, respectively. On the other hand, copper doping resulted in a reduced superconducting transition temperature $T_{c}$ $\simeq$ 2--4 K of resistivity even though superconducting diamagnetic signals were not observed in Ba(Fe$_{1-x}$Cu$_{x}$)$_{2}$As$_{2}$ \cite{ni2,garitezi,piva}.

Then, a natural question arises whether transition-metal doping, specifically copper doping, of BaNi$_{2}$As$_{2}$ can suppress the triclinic transition temperature and result in enhanced superconductivity similar to that observed in the phosphorus doping of BaNi$_{2}$As$_{2}$. 
In this paper, we report the results of copper doping of BaNi$_{2}$As$_{2}$, which revealed striking similarity with phosphorus doping. 
In fact, the superconducting transition temperature $T_{c}$ increased from 0.6 to 2.5--3.2 K as a result of the triclinic-tetragonal phase transition at copper doping $x$ = 0.16 for Ba(Ni$_{1-x}$Cu$_{x}$)$_{2}$As$_{2}$. 
Strong-coupling type superconductivity was observed in the tetragonal phase. 
The enhanced superconductivity was accompanied by significant phonon softening, which was observed via the Debye term of specific heat.

Single crystals of Ba(Ni$_{1-x}$Cu$_{x}$)$_{2}$As$_{2}$ were grown in two ways.
A mixture with a ratio of Ba:NiAs:Cu:As = 1:$4(1-x)$:$4x$:$4x$ \cite{kudo1} or 1:$2(1-x)$:$2x$:$2x$ was placed in an alumina crucible and sealed in an evacuated quartz tube.
The former mixture was heated at 700$^{\circ}$C for 3 h, slowly heated to 1150$^{\circ}$C, and cooled from 1150 to 1000$^{\circ}$C at a rate of 2$^{\circ}$C/h, followed by furnace cooling \cite{kudo1}. 
The latter mixture was heated at 700$^{\circ}$C for 3 h and then at 1000$^{\circ}$C for 72 h, followed by cooling to room temperature over 24 h. 
In both cases, single crystals with a typical dimension of 0.5--1.0 $\times$ 0.5--1.0 $\times$ 0.05 mm$^{3}$ were obtained. 
The results of powder x-ray diffraction, performed using a Rigaku RINT-TTR III x-ray diffractometer with Cu$K_{\alpha}$ radiation, showed that all specimens were in a single phase. 
Energy dispersive x-ray spectrometry (EDS) was used to determine the copper content $x$.
Samples with 0.485 $<$ $x$ $<$ 0.939 were rarely obtained, and attempts to synthesize Ba(Ni$_{1-x}$Cu$_{x}$)$_{2}$As$_{2}$ in this doping range resulted in phase separation between $x$ = 0.485 and 0.939, indicating a miscibility gap of BaNi$_{2}$As$_{2}$--BaCu$_{2}$As$_{2}$ solid solution. 
The samples were treated in a glove box filled with dried argon gas. 
The magnetization $M$ was measured using a Quantum Design MPMS. 
The electrical resistivity $\rho_{ab}$ (parallel to the $ab$ plane) and specific heat $C$ were measured using a Quantum Design PPMS.

Figure 1 shows the temperature dependence of the electrical resistivity $\rho_{ab}$ for Ba(Ni$_{1-x}$Cu$_{x}$)$_{2}$As$_{2}$. 
As previously reported, pure BaNi$_{2}$As$_{2}$ exhibits a transition at 130 K with a thermal hysteresis accompanying a sudden increase in resistivity upon cooling \cite{ronning,sefat,chen,kudo1}. 
For $x$ = 0.156 copper doping, the transition was significantly suppressed to 40 K, and the resistivity jump was small and broad. 
The transition appeared to be absent for $x$ = 0.292. 
These results suggested suppression of the triclinic phase at $x$ $\simeq$ 0.16.

\begin{figure}[t]
\begin{center}
\includegraphics[width=6cm]{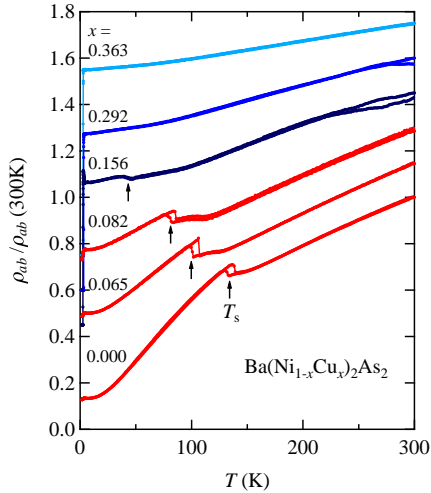}
\caption{
(Color online) Temperature dependence of the electrical resistivity parallel to the $ab$ plane, $\rho_{ab}$, normalized by the value at 300 K for Ba(Ni$_{1-x}$Cu$_{x}$)$_{2}$As$_{2}$. The data measured upon heating and cooling are plotted. For clarity, $\rho_{ab}/\rho_{ab}(300 {\rm K}$) is shifted by 0.15 with respect to all data. $T_{\rm s}$ is the phase transition temperature at which the tetragonal-to-triclinic phase transition takes place. 
}
\end{center}
\end{figure}

We observed that superconductivity with $T_{c}$ = 2.5--3.2 K emerged as soon as the triclinic phase was suppressed with copper doping, while $T_{c}$ 
$<$ 1.8 K (the lowest temperature measured) in the triclinic phase at $x$ $<$ 0.15. 
This behavior is demonstrated by the low-temperature magnetic susceptibility and resistivity data shown in Fig.~2. 
For $x$ = 0.292, the bulk superconductivity was evident from the full-shielding diamagnetic signal and sharp resistivity transition at 3.2 K.
On the other hand, the resistivity transition was broad for $x$ = 0.156, which occurred at the critical phase boundary of the triclinic and tetragonal phases. 
Superconductivity persisted up to $x$ = 0.363, while $T_{c}$ decreased slightly to 2.75 K. 
Superconductivity was not observed for $x$ = 0.485, which was the solubility limit of Cu for Ni. 
The low-temperature specific-heat data, shown in Fig.~2(c) and 3, provided further evidence of the enhanced superconductivity in the tetragonal phase. 
Pure BaNi$_2$As$_2$ exhibits a specific-heat jump at 0.6 K, as reported previously \cite{kudo1,ronning,sefat,kurita}. 
In the tetragonal phase at $x$ = 0.292, the specific-heat jump appeared at an elevated temperature of 3.2 K, in a consistent manner with the magnetic susceptibility and resistivity data. 
The specific-heat jump was significantly broadened for $x$ = 0.363, suggesting a reduced superconducting volume fraction and the disappearance of superconductivity before reaching the solubility limit at $x$ = 0.485.

\begin{figure}[t]
\begin{center}
\includegraphics[width=6.5cm]{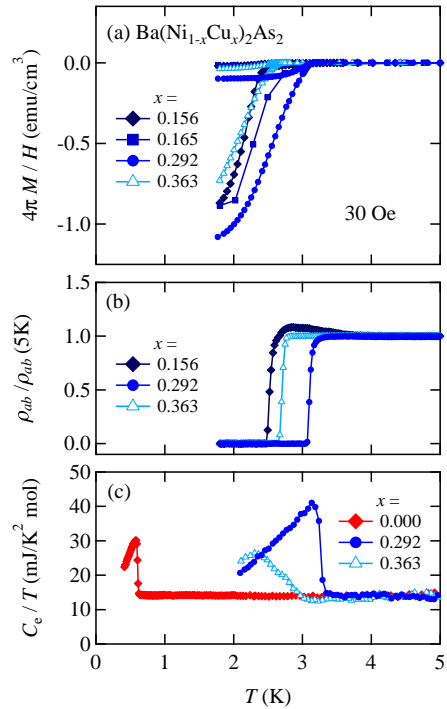}
\caption{
(Color online) 
(a) Temperature dependence of dc magnetization $M$ measured in a magnetic field $H$ of 30 Oe for Ba(Ni$_{1-x}$Cu$_{x}$)$_{2}$As$_{2}$ under zero-field cooling and field cooling. 
(b) Temperature dependence of the electrical resistivity parallel to the $ab$ plane, $\rho_{ab}$, normalized by the value at 5 K for Ba(Ni$_{1-x}$Cu$_{x}$)$_{2}$As$_{2}$. 
(c) Temperature dependence of the electronic specific heat divided by the temperature, $C_e / T$, for Ba(Ni$_{1-x}$Cu$_{x}$)$_{2}$As$_{2}$. $C_e$ was determined by subtracting the phonon contribution  $\beta T^3$ from the total specific heat $C$, as shown in Fig.~3. 
The specific-heat data for BaNi$_{2}$As$_{2}$ ($x$ = 0.0) are taken from Ref. \cite{kudo1}. 
}
\end{center}
\end{figure}

\begin{figure}[t]
\begin{center}
\includegraphics[width=6.5cm]{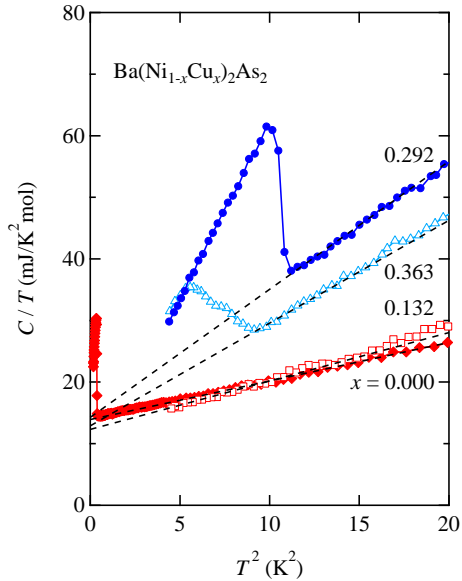}
\caption{
(Color online) The specific heat divided by the temperature, $C/T$, as a function of $T^{2}$ for Ba(Ni$_{1-x}$Cu$_{x}$)$_{2}$As$_{2}$. The dashed lines denote fits by $C/T = \gamma + \beta T^{2}$, where $\gamma$ is the electronic specific-heat coefficient and $\beta$ is a constant corresponding to the Debye phonon contributions. 
The specific-heat data for BaNi$_{2}$As$_{2}$ ($x$ = 0.0) are taken from Ref. \cite{kudo1}. 
}
\end{center}
\end{figure}

Figure 3 shows the specific heat divided by the temperature $C/T$ as a function of the squared temperature $T^{2}$. 
The normal-state data above $T_{c}$ could be well fitted by $C/T = \gamma + \beta T^{2}$, where $\gamma$ was the electronic specific-heat coefficient and $\beta$ was the coefficient of phonon contributions from which the Debye temperature $\Theta_{\rm D}$ was estimated. 
As shown in Fig.~3, the slope of the $C/T$ vs $T^{2}$ lines was almost unchanged in the triclinic phase for $x$ $<$ 0.16. 
The slope increased drastically for $x$ = 0.292 copper doping, suggesting significant phonon softening in the tetragonal phase. 
The slope decreased upon further copper doping in the tetragonal phase. 
Figure 4(c) shows the estimated Debye temperature $\Theta_{\rm D}$ as a function of the copper content $x$. 
In particular, $\Theta_{\rm D}$ showed significant reduction from 250 K for $x$ = 0.0 to 170 K for $x$ = 0.292.

In contrast to the strong dependence of both $\Theta_{\rm D}$ and $T_{c}$ on doping, the electronic specific-heat coefficient $\gamma$ (= 14 mJ/mol K$^{2}$) was almost independent of the copper content $x$, as shown from the almost unchanged intercept of the $C/T$ vs $T^{2}$ lines along the $C/T$ axis in Fig.~3. 
These observations suggested that the structural phase transition, as well as the enhanced superconductivity in the tetragonal phase, originated in the enhanced electron-phonon coupling due to soft phonons \cite{kudo1}. 
Indeed, the normalized specific-heat jump $\Delta C / \gamma T_{c}$ $\simeq$ 1.3 for pure BaNi$_2$As$_2$ in the triclinic phase, determined from the data shown in Fig.~2(c), was comparable to the value of the weak-coupling limit (= 1.43), while $\Delta C / \gamma T_{c}$ was enhanced in the tetragonal phase with a value of 1.9 at $x$ = 0.292, indicating strong-coupling superconductivity.

\begin{figure}[t]
\begin{center}
\includegraphics[width=6.5cm]{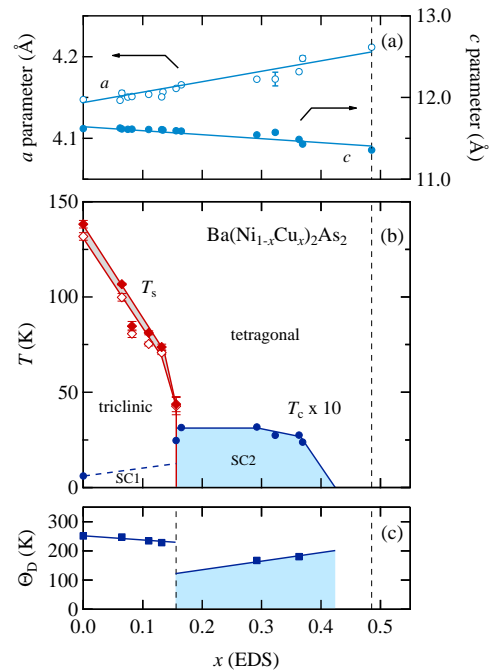}
\caption{
(Color online) 
(a) Lattice parameters $a$ and $c$ as a function of copper content $x$ at 300 K for Ba(Ni$_{1-x}$Cu$_{x}$)$_{2}$As$_{2}$. 
(b)  Electronic phase diagram of Ba(Ni$_{1-x}$Cu$_{x}$)$_{2}$As$_{2}$. 
The (blue) closed circles represent the superconducting transition temperatures $T_{c}$. For clarity, the values of $T_{c}$ are scaled by a factor of 10. SC1 and SC2 denote the superconducting phases. The (red) open and closed diamonds represent the tetragonal-to-triclinic structural transition temperatures $T_{\rm s}$ upon cooling and heating, respectively. 
(c) The Debye temperature $\Theta_{\rm D}$  as a function of phosphorous content $x$ for Ba(Ni$_{1-x}$Cu$_{x}$)$_{2}$As$_{2}$. $\Theta_{\rm D}$ is determined from the slope of the $C/T$ vs $T^{2}$ curves in Fig.~3.
}
\end{center}
\end{figure}

Our observations in Ba(Ni$_{1-x}$Cu$_{x}$)$_{2}$As$_{2}$ are summarized in the electronic phase diagram shown in Fig.~4(b). 
The triclinic phase transition temperature in pure BaNi$_{2}$As$_{2}$ was gradually suppressed with copper doping. 
As a result, the superconducting transition temperature was enhanced from $T_{c}$ = 0.6 K in pure BaNi$_{2}$As$_{2}$ to 2.5--3.2 K in the tetragonal phase at $x$ $>$ 0.16. 
The strong-coupling enhanced superconductivity was accompanied by a pronounced phonon softening that was induced by copper doping, while the electronic density of state at the Fermi level remained almost unchanged. 
All of these observations in copper-doped Ba(Ni$_{1-x}$Cu$_{x}$)$_{2}$As$_{2}$ were strikingly similar to those observed in phosphorus-doped BaNi$_{2}$(As$_{1-x}$P$_{x}$)$_{2}$ \cite{kudo1}. 
For BaNi$_{2}$(As$_{1-x}$P$_{x}$)$_{2}$, the triclinic phase transition temperature was suppressed with phosphorus doping, and the superconducting transition temperature was enhanced to 3.2--3.3 K in the tetragonal phase at $x$ $>$ 0.067. 
The electronic specific-heat $\gamma$ was almost unchanged upon phosphorous doping, and the Debye temperature $\Theta_{\rm D}$ exhibited significant reduction from 250 to 150 K at the structural phase boundary of $x$ = 0.067.

Dissimilarity between copper and phosphorus doping could be observed in the opposite doping dependence of lattice parameters. 
For Ba(Ni$_{1-x}$Cu$_{x}$)$_{2}$As$_{2}$, as shown in Fig.~4(a), the $a$ parameter increased with copper doping $x$, while $c$ decreased with ratios of $(1/a)(da/dx) = + 0.032$ and $(1/c)(dc/dx) = - 0.046$, which resulted in a volume increase of $(1/V)(dV/dx) = + 0.017$. 
In contrast, phosphorus doping resulted in $(1/a)(da/dx) = - 0.043$, $(1/c)(dc/dx) = + 0.013$, and $(1/V)(dV/dx) = - 0.072$ for BaNi$_{2}$(As$_{1-x}$P$_{x}$)$_{2}$ \cite{kudo1}, as expected from the small ionic radius of phosphorus and resultant chemical pressure. 
Thus, the reduction of the triclinic structural transition temperature upon copper/phosphorus doping could not be ascribed to the volume effect. 
This behavior was consistent with the weak effect of hydrostatic pressure on both the triclinic and superconducting transition temperatures \cite{park}.

Finally, we discuss the results of further doping. 
A miscibility gap of the copper content was observed between $x$ = 0.485 and 0.939 for Ba(Ni$_{1-x}$Cu$_{x}$)$_{2}$As$_{2}$. 
Superconductivity was absent when nickel was completely replaced by copper in BaCu$_{2}$As$_{2}$ \cite{saparov}. 
It has been reported that Cu $3d$ orbitals are fully occupied for BaCu$_{2}$As$_{2}$ \cite{wu}. 
Similarly, a miscibility gap of the phosphorus content existed above $x$ = 0.13 for BaNi$_{2}$(As$_{1-x}$P$_{x}$)$_{2}$ \cite{kudo1}. 
Superconductivity at $T_{c}$ = 2.5 K appeared when arsenic was completely replaced by phosphorus in BaNi$_{2}$P$_{2}$ \cite{mine,tomioka,kudo2,hirai}, but it was weak-coupling type superconductivity as demonstrated in BaNi$_{2}$(Ge$_{1-x}$P$_{x}$)$_{2}$ \cite{hirai}.

In conclusion, our studies demonstrated that enhanced superconductivity associated with phonon softening and the subsequent structural phase transition occurred upon copper doping of BaNi$_{2}$As$_{2}$. Specifically, the increase in $T_{c}$ from 0.6 to 2.5--3.2 K was related to the giant phonon softening of $\sim$50{\%} at the triclinic-to-tetragonal phase transition induced by copper doping. 
These features were strikingly similar to those observed in phosphorus doping of BaNi$_{2}$As$_{2}$ \cite{kudo1}. 
The similarity of copper and phosphorus doping of the nickel-based superconductor BaNi$_{2}$As$_{2}$ and dissimilarity of copper and phosphorus doping of the iron-based superconductor BaFe$_{2}$As$_{2}$ could provide further understanding of the role of chemical substitution in realizing high-temperature superconductivity. 

Part of this work was performed at the Advanced Science Research Center, Okayama University.
This work was partially supported by Grants-in-Aid for Scientific Research (No. 26287082, 15H01047, 15H05886, and 16K05451) provided by the Japan Society for the Promotion of Science (JSPS) 
and the Program for Advancing Strategic International Networks to Accelerate the Circulation of Talented Researchers from JSPS.

\end{document}